\documentclass[
 reprint,
superscriptaddress,
bibnotes,
 amsmath,amssymb,
 aps,prl,
floatfix,
]{revtex4-2}

\usepackage{graphicx}% Include figure files
\usepackage{dcolumn}% Align table columns on decimal point
\usepackage{bm}% bold math
\usepackage{braket}
\begin{document}

\title{State Readout of a Trapped Ion Qubit Using a Trap-Integrated Superconducting Photon Detector}

\author{S. L. Todaro}
\altaffiliation[Current address: ]{Research Laboratory for Electronics, Massachusetts Institute for Technology, Cambridge, Massachusetts 02139, USA}
\affiliation{Time and Frequency Division, National Institute of Standards and Technology, 325 Broadway, Boulder, Colorado 80305, USA}
\affiliation{Department of Physics, University of Colorado, Boulder, Colorado 80309, USA}
\author{V. B. Verma}
\affiliation{Applied Physics Division, National Institute of Standards and Technology, 325 Broadway, Boulder, Colorado 80305, USA}
\author{K. C. McCormick}
\altaffiliation[Current address: ]{Department of Physics, University of Washington, Seattle, Washington 98195, USA}
\affiliation{Time and Frequency Division, National Institute of Standards and Technology, 325 Broadway, Boulder, Colorado 80305, USA}
\affiliation{Department of Physics, University of Colorado, Boulder, Colorado 80309, USA}
\author{D. T. C. Allcock}
\affiliation{Time and Frequency Division, National Institute of Standards and Technology, 325 Broadway, Boulder, Colorado 80305, USA}
\affiliation{Department of Physics, University of Colorado, Boulder, Colorado 80309, USA}
\affiliation{Department of Physics, University of Oregon, Eugene, Oregon 97403, USA}
\author{R. P. Mirin}
\affiliation{Applied Physics Division, National Institute of Standards and Technology, 325 Broadway, Boulder, Colorado 80305, USA}
\author{D. J. Wineland}
\affiliation{Time and Frequency Division, National Institute of Standards and Technology, 325 Broadway, Boulder, Colorado 80305, USA}
\affiliation{Department of Physics, University of Colorado, Boulder, Colorado 80309, USA}
\affiliation{Department of Physics, University of Oregon, Eugene, Oregon 97403, USA}
\author{S. W. Nam}
\affiliation{Applied Physics Division, National Institute of Standards and Technology, 325 Broadway, Boulder, Colorado 80305, USA}
\author{A. C. Wilson}
\affiliation{Time and Frequency Division, National Institute of Standards and Technology, 325 Broadway, Boulder, Colorado 80305, USA}
\author{D. Leibfried}
\affiliation{Time and Frequency Division, National Institute of Standards and Technology, 325 Broadway, Boulder, Colorado 80305, USA}
\author{D. H. Slichter}
\email{daniel.slichter@nist.gov}
\affiliation{Time and Frequency Division, National Institute of Standards and Technology, 325 Broadway, Boulder, Colorado 80305, USA}

\date{\today}

\begin{abstract}

We report high-fidelity state readout of a trapped ion qubit using a trap-integrated photon detector.  We determine the hyperfine qubit state of a single $^9$Be$^+$ ion held in a surface-electrode rf ion trap by counting state-dependent ion fluorescence photons with a superconducting nanowire single-photon detector fabricated into the trap structure.  The average readout fidelity is 0.9991(1), with a mean readout duration of 46 $\mu$s, and is limited by the polarization impurity of the readout laser beam and by off-resonant optical pumping.  Because there are no intervening optical elements between the ion and the detector, we can use the ion fluorescence as a self-calibrated photon source to determine the detector quantum efficiency and its dependence on photon incidence angle and polarization.
\end{abstract}

\maketitle

Qubit state readout is an essential part of quantum computing and simulation~\cite{DiVincenzo2000, Nielsen2000}, including most quantum error correction protocols~\cite{Preskill1998, Knill2005}.  Trapped ion qubits are typically read out by driving an optical cycling transition with laser light and observing the presence or absence of ion fluorescence~\cite{Dehmelt1982}.  A fraction of the fluorescence photons from the ion are collected, usually with an objective, and imaged onto a photon-counting detector or camera; the number of photons counted over the duration of the readout process indicates the projected state of the qubit.  In general, counting just a few percent of the total fluorescence photons from the ion is sufficient to provide readout fidelities in excess of 0.99~\cite{Wineland1998}, and readout fidelities at or approaching 0.9999 have been reported~\cite{Myerson2008, Burrell2010, Christensen2019, Zhukas2020}.  Trapped-ion readout can also be accomplished using state-dependent interactions with a second ion followed by fluorescence readout of that ion, as in quantum logic spectroscopy~\cite{Schmidt2005}.

Increasing the number of qubits in trapped ion quantum processors and simulators can boost computational power, but presents the challenge of reading out the individual states of multiple ions in parallel.  One solution is to employ spatially-resolved detection, where each ion's fluorescence is ideally imaged onto a separate active detector region.  Fluorescence crosstalk, where photons from one ion are counted by a detector region dedicated to a different ion, can be tolerated to some degree before the readout fidelity is degraded~\cite{Burrell2010, Schindler2013b, Debnath2016}.  Alternatively, multi-ion readout can be achieved without spatially-resolved detection through time-domain-multiplexed illumination of individual ions, for example by separating ions into different locations in the trap and reading them out in series~\cite{Wan2019}.  This increases the duration of readout in proportion to the number of qubits, limiting utility for many-ion systems.  

A number of groups use microfabricated surface-electrode traps~\cite{Seidelin2006a}, which can hold many ions and feature complex designs with multiple trapping zones~\cite{Amini2010, Guise2015, Maunz2016, Pino2020}, as a path toward large-scale trapped ion quantum computing.  The separate trapping zones can be used for different algorithmic tasks such as memory, readout, or gate operations~\cite{Wineland1998,Kielpinski2002, Pino2020}.  A natural method for simultaneous readout in such traps is to integrate on-chip photon collection features into the readout zones, such as optical fibers~\cite{VanDevender2010}, high-numerical-aperture (NA) micro-optics~\cite{Merrill2011a, Clark2014, Ghadimi2016}, or high-reflectivity trap surfaces~\cite{Herskind2011, VanRynbach2016}.  However, these solutions all rely on separate photon detectors or cameras, and some still require external objectives made with bulk optics.  Alternatively, spatially-resolved detectors fabricated directly into a surface-electrode trap could perform parallel qubit readout without external collection optics or detectors, with readout signals coupled out of the trap chip as electrical pulses~\cite{Leibfried2007, Eltony2013, Slichter2017}.  Such a readout architecture frees up the space and optical access used by bulk optics objectives and cameras, and potentially enables surface-electrode traps to be tiled in the third dimension, especially when combined with integrated photonics for light delivery~\cite{Mehta2016, Niffenegger2020, Mehta2020}.  It also eliminates the need for imaging system alignment, and can in principle be scaled to ion traps with many trap zones.  

In this Letter, we report the first use of a trap-integrated photon detector for high-fidelity state readout of an ion qubit.  We use a superconducting nanowire single-photon detector (SNSPD) co-fabricated with a surface-electrode ion trap to detect fluorescence photons at 313\,\,nm from a single $^9$Be$^+$ ion, achieving qubit state readout with fidelity 0.9991(1) in an average of 46\,\,$\mu$s using an adaptive Bayesian readout scheme~\cite{Hume2007, Myerson2008}.  Using the ion as a tunable, self-calibrating source of photons with known flux and polarization, we characterize the detection efficiency of the SNSPD as a function of incidence angle and polarization, finding agreement with theoretically predicted values.  We also study the effect of the trapping rf fields on the SNSPD perfomance, and characterize motional heating of an ion confined over the SNSPD.  

SNSPDs are a class of photon detectors with high quantum efficiency~\cite{Marsili2012, Reddy2019}, low dark counts~\cite{Wollman2017, Hochberg2019}, and picosecond timing jitter~\cite{Korzh2018}.  Recent experiments have shown quantum efficiencies in the UV of 75~\% to 85~\% at operating temperatures up to 4 K, a parameter regime relevant for ion trap applications~\cite{Slichter2017, Wollman2017}.  Ion fluorescence photons collected with traditional high-NA bulk optics have been counted by a fiber-coupled SNSPD in a stand-alone cryostat to perform fast, high-fidelity qubit readout~\cite{Crain2019}.  However, surface-electrode ion traps present a challenging electromagnetic and thermal environment for integrated SNSPDs: SNSPDs requiring low-noise bias currents of a few microamps must be placed close to trap electrodes with rf potentials of tens to hundreds of volts oscillating at up to {$\sim100$~MHz}.  Furthermore, the superconducting transition temperature $T_c$ of the SNSPD should be at least $\sim 25$~\% higher than the temperature at the surface of the trap (typically $\gtrsim4$ K) to achieve the best detection efficiency~\cite{Verma2014, Wollman2017, Engel2013}.  Combining the separate microfabrication processes for SNSPDs and ion traps while maintaining high device yield and good performance is also a challenge~\cite{supp}.  However, previous work has demonstrated successful integration and operation of SNSPDs on a test chip simulating the thermal and electromagnetic environment of an ion trap~\cite{Slichter2017}.

\begin{figure}[tb]
\includegraphics[width=0.45\textwidth]{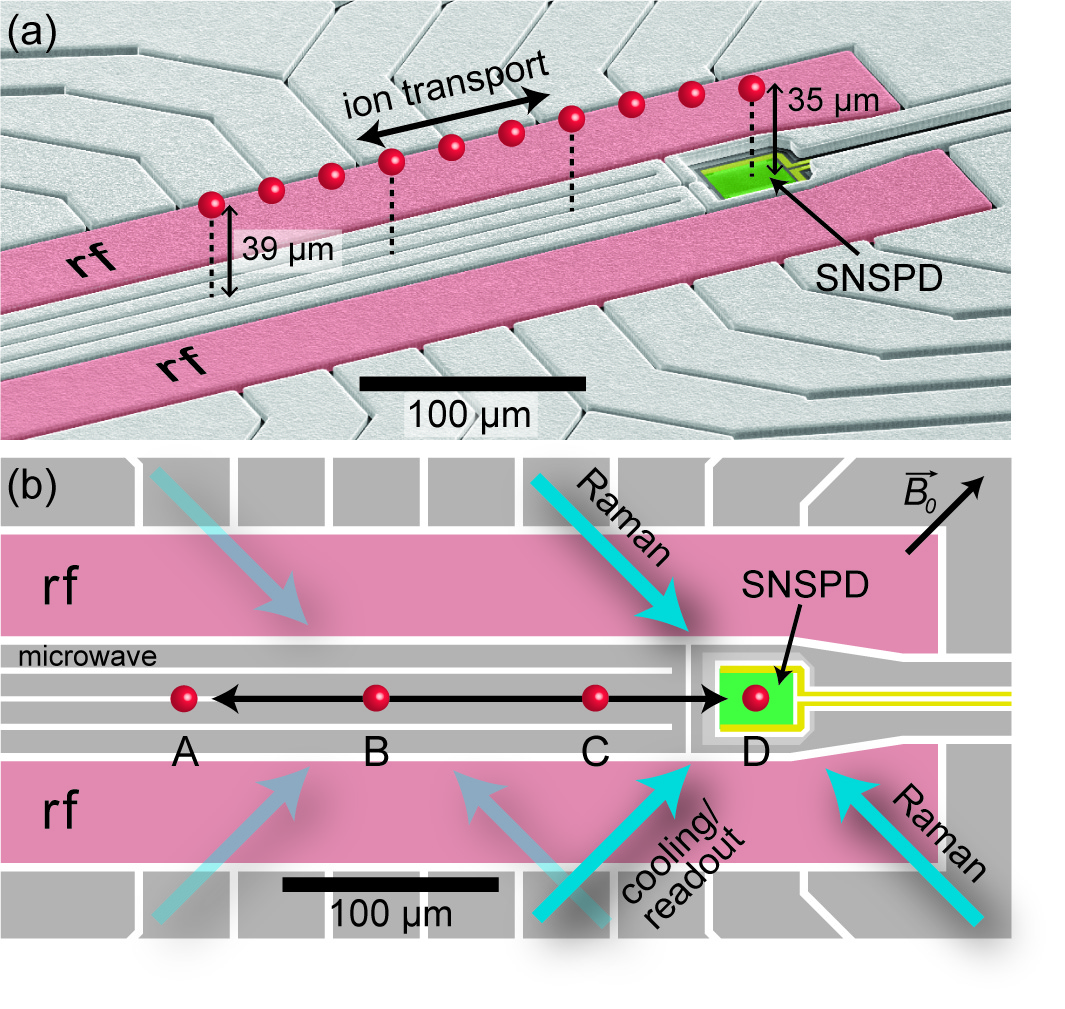}
\centering
\caption{Trap configuration.  (a) False-color scanning electron micrograph of the ion trap showing the rf electrodes (pink), SNSPD (green), and SNSPD bias leads (yellow).  A trapped ion (red sphere, shown in multiple positions along the rf null line) can be transported along the trap axis by applying appropriate time-varying potentials to the outer segmented electrodes (grey).  (b) Top view scale diagram showing four labeled trapping zones A-D along the trap axis (double-headed black arrow), as well as the geometry of the laser beams (blue solid arrows, here shown directed at zone D) and quantization magnetic field $\vec{B}_0$, which all lie in the plane of the trap at 45$^\circ$ angles to the trap axis.  The laser beams can be translated horizontally to follow the ion as it is transported between zones, as indicated by the faint laser beam arrows directed at zone B.}
\label{fig:setup}
\end{figure} 

The trap used in this work, shown in false color in Figure~\ref{fig:setup}, is a linear rf (Paul) surface-electrode trap with an SNSPD (green) fabricated on the trap substrate.  The rf electrodes (pink) provide confinement transverse to the trap axis (shown as a double-headed black arrow in Fig.~\ref{fig:setup}(b)), while the surrounding segmented electrodes (grey) confine the ion at adjustable positions along the rf null line, from directly over the SNSPD (zone D) to 264 $\mu$m away from the SNSPD center (zone A).  The ion is held ${\approx\,39\,\mu}$m above the top surface plane of the trap electrodes, dropping by design to a smaller distance of ${\approx\,29\,\mu}$m above this plane when centered over the SNSPD, which is recessed another 6 $\mu$m below this plane.  When the ion is in zone D, this gives an effective NA of 0.32 for the SNSPD; accounting for the dipole emission pattern of the ion fluorescence, 2.0(1)~\% of the emitted photons will strike the SNSPD active region~\cite{supp}.  An integrated current-carrying electrode running along the length of the trap between the rf electrodes generates microwave-frequency magnetic fields for qubit control.  The trap electrodes are made of electroplated Au on an intrinsic Si substrate, while the SNSPD is made of amorphous Mo$_{0.75}$Si$_{0.25}$, and has a superconducting transition temperature of 5.2 K~\cite{supp}.  The trap is installed in an ultra-high-vacuum low-vibration closed-cycle cryostat operated at a temperature of $\approx3.5$ K~\cite{Todaro2020}.  

We trap a single $^{9}$Be$^+$ ion with typical motional frequencies of $\sim 2$ MHz in the axial direction and 5~MHz to 10~MHz in the radial directions (normal to the trap axis).  The potential on the trap rf electrodes has a peak amplitude of 8.8 V at a frequency of 67.03 MHz.  A magnetic field of 0.56 mT, in the plane of the trap electrodes and oriented at 45$^\circ$ relative to the trap axis (see Fig.~\ref{fig:setup}(b)), lifts the degeneracy between hyperfine sublevels and defines the quantization axis.  This field had no discernible effect on SNSPD performance, consistent with other studies at higher fields~\cite{Korneev2014, Korneeva2020, Polakovic2020}.  We use the $\ket{F=2,m_F=-2}\equiv\ket{\downarrow}$ and $\ket{F=1, m_F=-1}\equiv\ket{\uparrow}$ states within the $2s\, ^2S_{1/2}$ hyperfine manifold as our qubit, which has a transition frequency of ${\omega_0/2\pi \approx 1.260}$\,GHz.  We prepare $\ket{\downarrow}$ by optical pumping on the ${2s\, ^2S_{1/2}\leftrightarrow2s\, ^2P_{3/2}}$ transitions at 313 nm with $\sigma^-$ polarized light. The qubit is read out by detecting fluorescence from the laser-driven $\ket{\downarrow}\leftrightarrow\ket{2s\, ^2P_{3/2}, F=3, m_F=-3}$ cycling transition. Before detection, microwave current pulses on the trap-integrated microwave electrode are used to transfer (``shelve''~\cite{Dehmelt1982}) population from $\ket{\uparrow}$ to the $\ket{aux}\equiv\ket{2s\, ^2S_{1/2}, F=1, m_F=1}$ state for improved readout fidelity.  A pair of counterpropagating laser beams detuned 80 GHz blue of the $2s\, ^2S_{1/2}\leftrightarrow2s\, ^2P_{1/2}$ transition at {313~nm} are used to drive stimulated Raman transitions on the first order secular motional sidebands, enabling sideband cooling and motional heating rate measurements~\cite{Monroe1995a}.  

\begin{figure}[tb]
\includegraphics[width=0.48\textwidth]{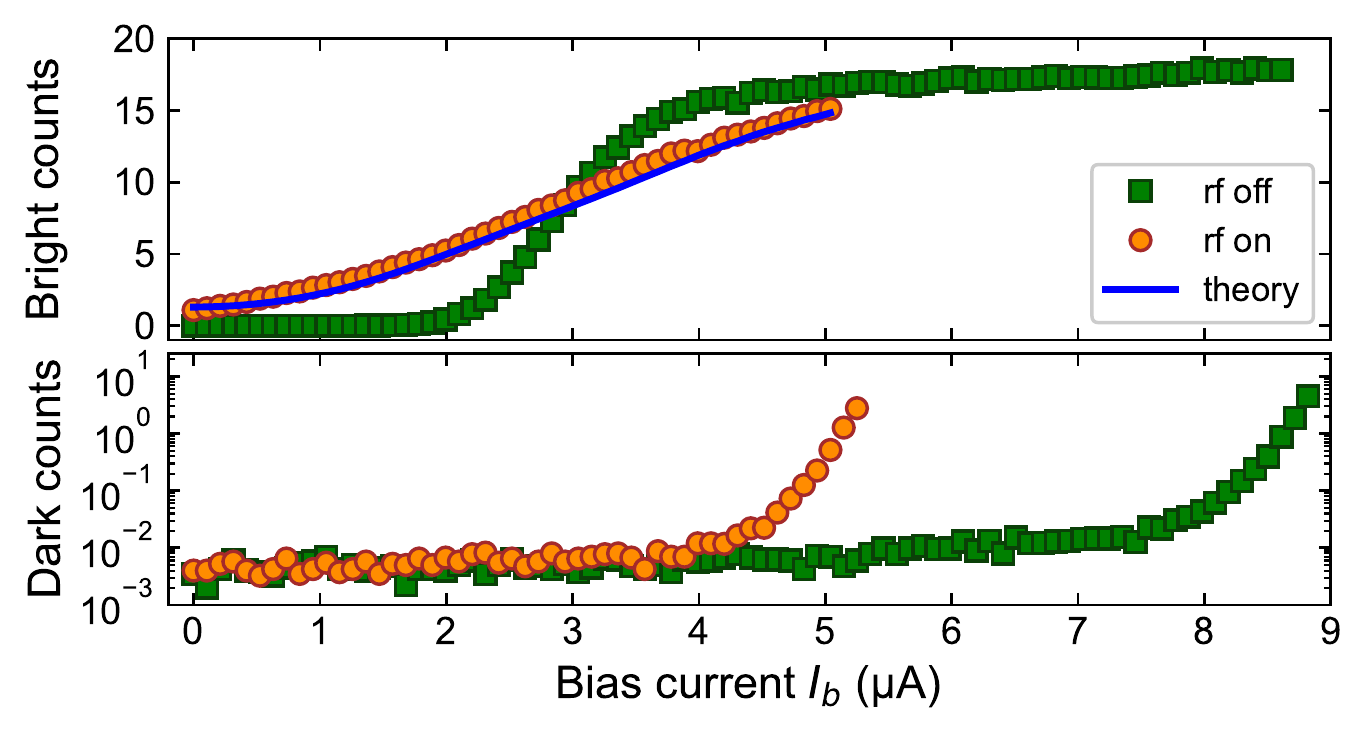}
\centering
\caption{Impact of trap rf on SNSPD performance.  We plot bright (top) and dark (bottom, log scale) counts in a 200 $\mu$s detection window versus SNSPD bias current, with trap rf either off (green squares) or on (orange circles), using laser scatter to simulate ion fluorescence for the bright counts.  The blue line is a fit to a theoretical model accounting for induced rf currents in the SNSPD.  The bright counts are background-corrected by subtracting the measured dark counts at each bias current.  The 68 \% confidence intervals on the reported values are smaller than the plot symbols.}
\label{fig:ibias}
\end{figure} 

One terminal of the SNSPD is grounded close to the trap chip, while the other is connected via a 50 $\Omega$ coaxial cable to room temperature bias and readout electronics~\cite{supp}.  The SNSPD bias current is applied only during readout and is off at other times.  The output signal is amplified and filtered to remove parasitic pickup of the trap rf drive before being digitized by a high-speed Schmitt-trigger comparator.  The digital pulses are counted and timestamped with 1 ns resolution.  

The performance of the SNSPD at 3.45 K was evaluated with the trap rf both off and on.  Because an ion cannot be held without trap rf, these measurements were carried out using a simulated ion fluorescence signal generated by laser beam scatter.  The beam position and intensity were chosen to give SNSPD count rates similar to those from a single ion in the trap.  Figure \ref{fig:ibias} plots the bright counts (laser on) and dark counts (laser off) during a 200 $\mu$s detection window as a function of the applied SNSPD bias current $I_{b}$, both with and without trap rf.  The $I_b$ at which the critical current density of the superconducting nanowire is exceeded, known as the switching current, is $\approx8.9\,\mu$A.  The trap rf decreases the maximum dc bias current $I_{m}$ that can be applied without driving the SNSPD to the normal (non-superconducting) state.   We attribute this reduction to induced rf currents modulating the bias current of the SNSPD~\cite{Slichter2017, supp}; a two-parameter fit to a theoretical model for induced rf currents, shown as the blue line, agrees quantitatively with experimental data~\cite{supp}.  Despite the reduction in $I_{m}$, the maximum bright counts with the trap rf on are only 17~\% lower than the maximum bright counts with the rf off.  The mean dark counts per detection, both with and without rf, remain below $10^{-2}$ for $I_{b}$ at least $\sim$1 $\mu$A below the rf-dependent $I_{m}$.  We emphasize that the dark counts in Fig.~\ref{fig:ibias} are measured in the absence of laser light, and are due to residual stray room light or intrinsic detector dark counts~\cite{Yamashita2011}.  In the experiments described below, the dark count rate is dominated by stray laser light.  

Ion loading occurs in trap zone A, 264 $\mu$m from the SNSPD center, and the trapped ion is transported to the detector (zone D) using time-varying potentials on the segmented outer electrodes.  When the ion is held above the SNSPD, the detector count rates from ion fluorescence can be combined with knowledge of the excited state lifetime $1/\Gamma=8.850(2)$ ns~\cite{Safronova2013} of the ion and the ion-detector geometry (including the ion dipole radiation pattern) to provide an absolute calibration of the system detection efficiency (SDE) of the SNSPD.  The SDE is defined as the fraction of photons incident on the SNSPD that register as counts in the readout electronics.  We vary the intensity of the readout laser beam and fit the corresponding count rates to determine the count rate when the atomic fluorescence transition is driven with a saturation parameter $s\gg1$~\cite{Budker2008}.  The background count rate, arising from stray laser scatter not due to the ion, can be subtracted by preparing the ion in a non-fluorescing state and measuring the count rate.  Using this technique, we extract an SDE of 48(2) \% with the trap rf on and $I_b=4\,\mu$A; accounting for the effects of rf and $I_b<I_m$, this would correspond to a maximum SDE of 65(5)~\% without rf~\cite{supp}.  This number is slightly lower than the theoretical design SDE of 72 \% based on nanowire geometry~\cite{supp}. 

\begin{figure}[t]
\includegraphics[width=0.49\textwidth]{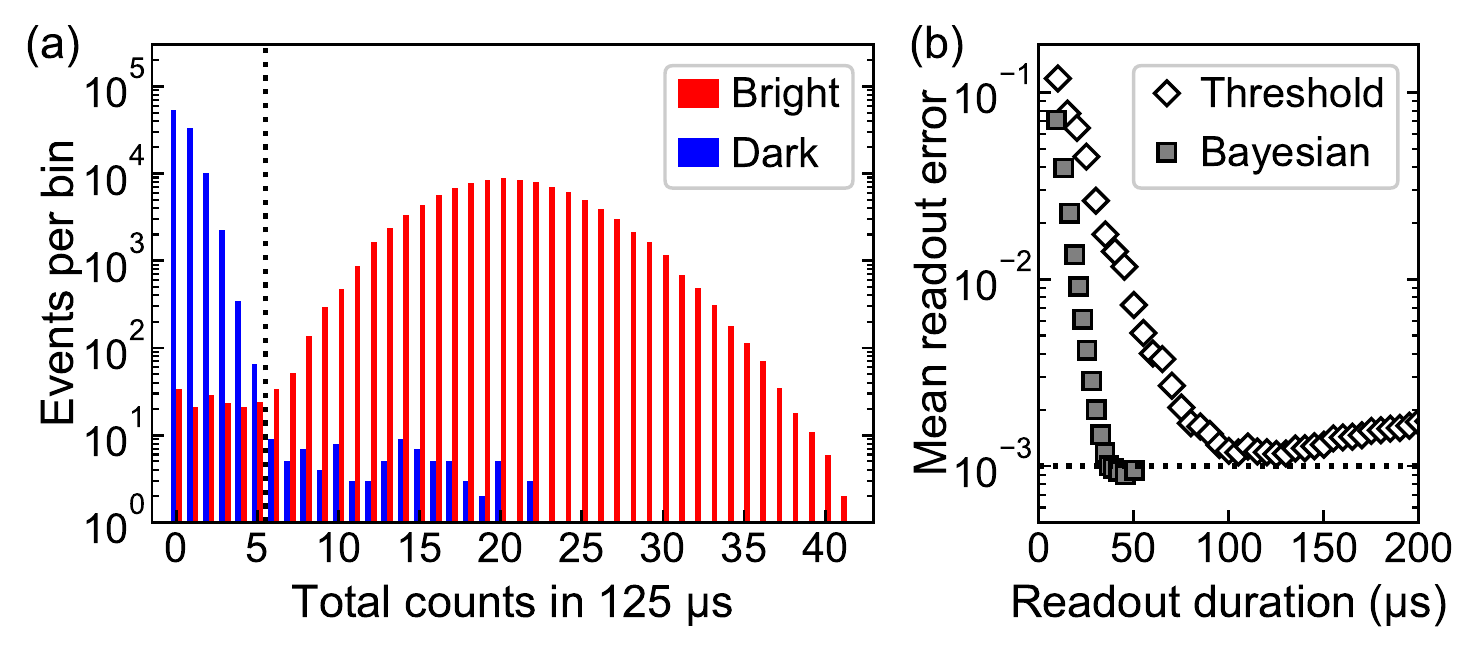}
\centering
\caption{Counts and readout error.  (a) Count histograms (log scale) for $10^5$ trials each of preparing the bright (red) and dark (blue) states, using a 125 $\mu$s detection window.  The dashed vertical line indicates the optimal threshold for state discrimination. (b) Mean readout error for $2\times10^5$ trials, half prepared dark and half prepared bright, using either standard thresholding or adaptive Bayesian methods for state determination.  For the Bayesian method, the horizontal axis is the mean readout duration before reaching a given state determination confidence level.  The dashed horizontal line indicates $10^{-3}$ mean readout error.  Statistical uncertainty in the mean readout error at the 68 \% confidence level is smaller than the plot symbols.}
\label{fig:hist}
\end{figure}

To characterize the fidelity of the qubit state readout, we prepare the ion in either the fluorescing ``bright'' $\ket{\downarrow}$ state or the shelved ``dark'' $\ket{aux}$ state and apply the readout laser beam for 500 $\mu$s.  We record the timestamps of all photons counted during this period, which enables us to vary the readout duration in post-processing.  We use heralding to improve the state preparation fidelity.  We define the first 50 $\mu$s of the data as the heralding period, and retain for further analysis only those trials with zero photon counts in this period as prepared in ``dark'', and those trials with eight or more photon counts as prepared in ``bright''.  This method reduces the contribution of state preparation error to the total measurement error.  We then analyze the readout fidelity for these trials, using only photon count data from after the heralding period, whose end defines the start of the readout period.  Figure {\ref{fig:hist}(a)} shows histograms of measured photon counts for both states using a readout duration of 125 $\mu$s after the heralding period, with a dotted line showing the threshold number of counts for optimal discrimination of bright and dark states~\cite{supp}.  The fidelity is limited by non-Poissonian tails that cross this threshold, arising from off-resonant pumping of $\ket{aux}$ into $\ket{\downarrow}$, and from imperfections in the $\ket{\downarrow}\leftrightarrow\ket{2s\,^2P_{3/2}, F=3, m_F=-3}$ cycling transition due to polarization impurity and trap-rf-induced state mixing.  The minimum readout error with the thresholding method is $1.2(1)\times10^{-3}$ at a readout duration of 125 $\mu$s.  We also analyze the measured state using a variant of the adaptive Bayesian method from Ref.~\cite{Myerson2008}; details are given in the supplemental material~\cite{supp}.  As shown in Fig. \ref{fig:hist}(b), the mean readout duration to reach a given error level is shorter than for the threshold method, and the minimum readout error of $9(1)\times10^{-4}$, achieved with an average readout duration of 46 $\mu$s, is smaller than can be achieved with thresholding. The corresponding maximum readout fidelities are 0.9988(1) and 0.9991(1) for the thresholding and Bayesian methods, respectively.  

The motional heating rate of the axial mode was measured in trap zone B, away from the SNSPD, to be 63(6) quanta/s at a frequency of $\omega/2\pi=2$ MHz, scaling with frequency as $\omega^{-1.7(7)}$.  When centered directly over the SNSPD in zone D, the axial mode heating rate was measured to be 113(14) quanta/s at $\omega/2\pi=5.3$ MHz.  Assuming heating rate distance scaling of $d^{-4}$~\cite{Turchette2000, Brownnutt2014} and the measured frequency scaling from zone B, the scaled electric field noise over the SNSPD is estimated to be roughly 6 times higher than that over the gold electrodes, but is still on par with state-of-the-art values reported in cryogenic ion traps~\cite{Brownnutt2014}.  It is unclear whether this increase is due to noise from the wideband SNSPD bias line, to materials properties of the SNSPD, or to some other mechanism.  

When the SNSPD outputs a pulse, some portion of the nanowire will stay at ground potential while the remainder will track the output voltage.  This causes a brief impulsive electric field ``kick'' to the ion, exciting its motion.  During readout, this effect can be neglected, as the ion temperature is determined primarily by the scattering of the resonant readout laser beam from the ion.  However, during operations such as stimulated Raman transitions when the ion does not spontaneously emit many photons, SNSPD pulses from stray laser light can become the dominant source of heating.  Even when the bias current is off, the SNSPD will occasionally pulse in response to photons when the trap rf is on, as seen in Fig. \ref{fig:ibias}.  We measure the resulting heating rate on the 5.3 MHz axial mode in zone D to be 0.009(5) quanta per SNSPD count.  This effect limited our ability to perform Raman sideband cooling of an ion held over the SNSPD, due to stray light from the Raman laser beams.  In a large-scale processor, operations with high-power Raman beams could be carried out in other trap zones, with the ion(s) transported to the readout zone(s) afterward.  The addition of optically transparent SNSPD shielding electrodes may permit operations with high-power Raman beams to be performed in trap zones with integrated SNSPDs, while also reducing induced rf currents in the SNSPDs.  

\begin{figure}[tb]
\includegraphics[width=0.48\textwidth]{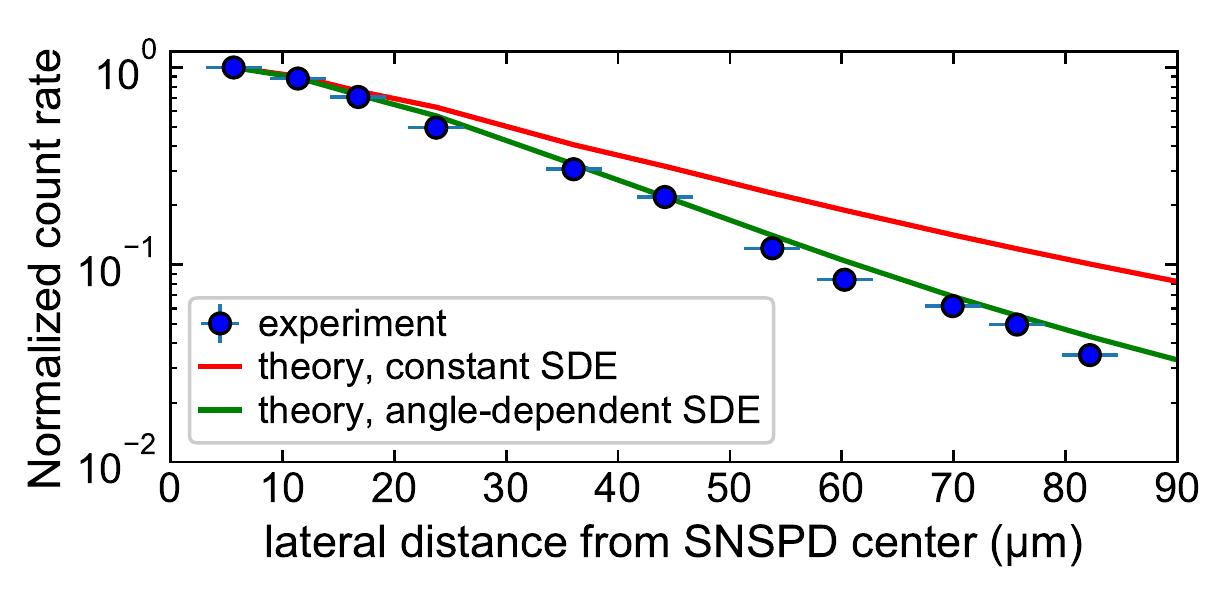}
\centering
\caption{Count rate spatial dependence.  Moving the ion along the trap axis away from the SNSPD reduces the count rate (blue circles) more strongly than is expected based on detector solid angle and ion dipole radiation pattern alone (red line). Including the calculated angular dependence of SNSPD SDE improves agreement (green line). The 68 \% confidence intervals on the count rates are smaller than the symbols; those on the theoretical calculations are narrower than the plotted lines.}
\label{fig:crosstalk}
\end{figure} 

Crosstalk from ions in neighboring readout zones will impact the fidelity of parallel readout with trap-integrated SNSPDs.  We characterized the crosstalk strength by measuring the SNSPD count rate as a function of the ion position along the trap axis.  In Fig. \ref{fig:crosstalk}, we plot the background-subtracted SNSPD count rate when the ion transition is driven with $s\gg 1$, normalized to the highest measured value, as a function of ion distance from the SNSPD center (zone D) along the trap axis~\cite{supp}.  The red curve shows the numerically calculated value assuming constant detector SDE, while the green curve uses a polarization- and incidence-angle-dependent SDE derived from finite element analysis of the SNSPD~\cite{supp}.  The improved agreement between the data and the angle-dependent SDE (versus constant SDE) provides, to the best of our knowledge, the first experimental measurement of the dependence of SNSPD SDE on photon incidence angle.  Both theory curves are normalized to the left-most experimental data point; this overall scaling accounts for experimental reductions in the SDE due to bias currents below $I_m$ and rf pickup~\cite{supp}.  The angle dependence of the SDE would help reduce crosstalk errors for parallel qubit readout below the level predicted simply from solid angle and dipole emission pattern considerations.  

Our results provide a path for scalable qubit readout in ion traps.  By combining multi-pixel SNSPD readout~\cite{Zhao2017, Wollman2019} with trap-integrated photonic waveguides for laser light delivery~\cite{Mehta2016, Niffenegger2020, Mehta2020}, it would be possible to create an ion trap without any free-space optical elements, potentially bringing substantial stability and performance improvements.  Finally, this work demonstrates the usefulness of individual trapped ions as well-characterized, tunable, high-precision photon sources for absolute calibration of single-photon detectors.  

We thank R. Srinivas and A. L. Collopy for a careful reading of the manuscript.  The device was fabricated in the Boulder Microfabrication Facility at NIST.  At the time the work was performed, S.L.T., K.C.M., and D.T.C.A. were Associates in the Professional Research Experience Program (PREP) operated jointly by NIST and the University of Colorado.  This work was supported by the NIST Quantum Information Program and IARPA.

S.L.T. and D.H.S. performed the experiments and analyzed the data; D.H.S. and V.B.V. designed and fabricated the trap chip; D.H.S. and S.L.T. designed and built the apparatus with contributions from K.C.M., A.C.W., D.L., and D.T.C.A.; D.L. proposed the integration of an SNSPD into an ion trap, with input from S.W.N., D.J.W., and R.P.M.; D.H.S. supervised the research and wrote the manuscript; A.C.W., D.L., and D.J.W. secured funding and provided additional supervision, assisted by S.W.N. and R.P.M.; all authors participated in experimental design, prototyping and testing efforts, and manuscript editing.

\bibliography{detector}

\begin{thebibliography}{68}
\bibitem [{dis()}]{disclaimer}%
  \BibitemOpen
  \href@noop {} {}\bibinfo {note} {Commercial products are identified in this
  paper for informational purposes only. Such identification does not imply
  recommendation or endorsement by the National Institute of Standards and
  Technology, nor is it intended to imply that the products identified are
  necessarily the best available for the purpose.}\BibitemShut {Stop}%
\end{thebibliography}

%%%%%%%%%% SUPPLEMENT %%%%%%%%%%%%%%

\newpage
\section{Supplemental Material}
\setcounter{equation}{0}
\setcounter{figure}{0}
\renewcommand{\theequation}{S\arabic{equation}}
\renewcommand{\thefigure}{S\arabic{figure}}

\subsection{Trap fabrication}

\begin{figure}[t]
\includegraphics[width=0.48\textwidth]{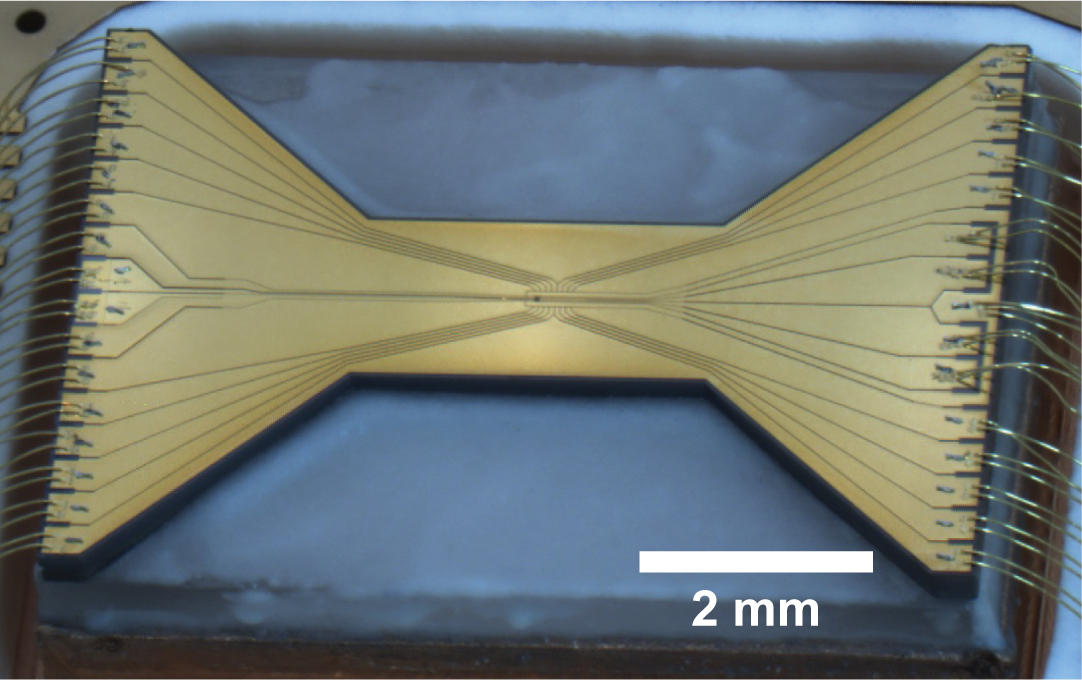}
\centering
\caption{Photograph of mounted trap. Note that the trap orientation is rotated by $180^\circ$ relative to Fig.~\ref{fig:setup}, with the detector here on the left.}
\label{fig:trap_overview}
\end{figure}

The trap is fabricated on an intrinsic Si substrate ($\rho>20$ k$\Omega$-cm). A 134 nm-thick layer of SiO$_2$ is deposited first by plasma-enhanced chemical vapor deposition (PECVD), followed by evaporation of 5 nm Ti and 50 nm Au (patterned with liftoff) to provide electrical contacts for the SNSPD and a seed layer for electroplating.  A second liftoff deposition of 350 nm of Au is added on the SNSPD leads only, to reduce their series resistance and thus prevent SNSPD latching~\cite{Annunziata2010, Kerman2013}.  This is necessary because the leads are much narrower than typical SNSPD leads, a design choice made to minimize capacitive coupling to the trap rf electrodes.  An 8 nm-thick layer of amorphous Mo$_{0.75}$Si$_{0.25}$, capped with 2 nm of amorphous Si, is deposited by dc magnetron sputtering.  This layer is then patterned with optical and electron beam lithography and inductively-coupled plasma reactive ion etching (ICP RIE) in an SF$_6$ plasma to form the nanowire meander~\cite{Slichter2017, Reddy2019}.  The meander covers an active area of $22\,\mu\mathrm{m}\, \times\,20\,\mu$m, with nanowires of 110 nm width on a 170 nm pitch, aligned with the trap axis.  The thickness of the SiO$_2$ and the dimensions of the nanowire are chosen using rigorous coupled-wave analysis (RCWA) simulations~\cite{Moharam1995} to optimize absorption of normally-incident 313 nm photons.  We calculate an estimated maximum SDE of 72~\% from RCWA when averaging over polarizations for normally incident photons, as described later in this Supplemental Material.  Following nanowire fabrication, 6 $\mu$m-thick Au trap electrodes are deposited by electroplating in a commercially available gold sulfite plating solution at $60^\circ$ C using a 10 $\mu$m-thick photoresist mask.  The patterned nanowire and its connecting lead electrodes are electrically and chemically isolated from the electroplating bath by the photoresist mask.  The Ti/Au seed layer in the gaps between electroplated electrodes is subsequently removed by Ar ion milling; again, the nanowire and its leads are protected by photoresist during this step.  Finally, the trap substrate is patterned in a ``bowtie'' shape~\cite{Maunz2016}, as seen in Fig.~\ref{fig:trap_overview}, by deep reactive ion etching (DRIE), with the active area of the trap on the 1.8 mm-wide central isthmus.  This shape makes it possible for focused laser beams to address the ion without significant clipping by the edge of the trap substrate.  The entire trap chip, including the nanowire, is protected by the resist used to define the bowtie shape.  Resist is stripped between fabrication steps using acetone.  The liftoff steps require a second strip in 1-methyl-2-pyrrolidone (NMP).

We observed that the fabrication steps following the nanowire patterning had a strong negative impact on SNSPD yield.  To study this, we measured the room-temperature resistance of every fabricated nanowire immediately after patterning, before the electroplating resist was applied, after the electroplating resist was stripped, after the ion milling resist was applied, after the ion milling resist was stripped, and after the DRIE resist was stripped.  The nanowire resistances range from 6~M$\Omega$ to 11~M$\Omega$ for typical functioning SNSPDs, depending on the SNSPD active area and nanowire width and pitch.  These resistances increased by roughly 12~\% to 30~\% over the course of the subsequent fabrication steps, suggesting either oxidation or etching of the superconducting film.  Larger resistance increases were strongly correlated with reduced nanowire switching currents or failure to superconduct at 3.6 K, even without trap rf.  The correlation was weaker for the smallest observed resistance increases.  For comparison, leaving the devices to sit unprotected in air for one month immediately following nanowire patterning caused a resistance increase of approximately 2.5~\%.  The dominant contribution to the measured nanowire resistance increase came between when the electroplating resist was applied and when it was stripped following electroplating.  The exact manner in which the electroplating process affects the nanowires is not known.   However, it is unlikely to be purely thermal, because the nanowire experiences higher temperatures for similar duration during the DRIE step, which has negligible impact on the nanowire resistance.  It also seems unlikely to be electrochemical, because the nanowire is isolated from the electroplating bath by the resist mask.  Applying and then stripping the resist mask, without electroplating, does not cause a substantial resistance increase.  

Future improvements might include coating the nanowire with a transparent dielectric layer such as SiO$_2$ immediately following patterning, which can provide an anti-reflection coating to increase SDE~\cite{Wollman2017, Reddy2019} and may improve nanowire robustness to subsequent fabrication steps.  For ion trapping applications, such a large area of exposed dielectric is often undesirable due to its tendency to accumulate stray charge, which can perturb the trapping potentials, so a transparent shielding electrode (either a continuous conducting film or a conducting mesh) would need to be fabricated on top of the dielectric.  This top conducting layer could also serve as part of an electrical shield for the SNSPD, reducing the induced rf currents in the SNSPD, the electric field ``kick'' at the ion due to SNSPD pulses, and any electric field noise at the ion originating from the meander or its bias and readout circuitry.  

\subsection{Experimental apparatus}

To aid in setup and calibration, and to enable detection of the qubit state when the ion is not near the SNSPD, a refractive objective outside the vacuum system with $\mathrm{NA}=0.38$ is used to image the trap or ion onto an electron-multiplied charge-coupled device (EMCCD) camera, or to collect fluorescence photons from the ion to be counted with a photomultiplier tube (PMT).  The objective was translated to image different regions of the trap as needed.  Photon counting with the PMT was carried out simultaneously with photon counting using the SNSPD, enabling photon arrival time correlations to be measured.  For an ion trapped in zone D, the mean bright (dark) photon count rate for the SNSPD was $\approx2.5$ ($\approx1.3$) times higher than for the PMT.  

All 313 nm laser beams are derived from frequency doubling of frequency-summed IR fiber lasers~\cite{Wilson2011c}.  Ions are loaded into the trap from a thermal flux of neutral Be atoms by resonance-enhanced two-photon photoionization~\cite{Burd2020} using a 235 nm laser beam derived from a frequency-quadrupled continuous wave Ti:sapphire laser.  The ion distance to the top surface plane of the trap electrodes (``ion height''), as well as its distance from zone D along the trap axis (``lateral distance''), is measured for each ion position by using motorized lenses to scan the readout laser beam across the ion.  The beam is scanned both parallel and perpendicular to the trap surface, and we fit to the observed photon counts on the PMT to determine the beam position for maximum fluorescence.  The ion height calibration is completed by scanning the beam toward the trap surface and performing a Gaussian fit to the PMT counts of the light scattered off the trap as the beam centerline approaches and then passes the plane of the trap electrode top surface.  This PMT count rate will be proportional to the total intensity in the thin slice of beam parallel to the trap surface that is being scattered by the surface.  Since scanning the beam position as described changes only the slice of beam being scattered in the objective field of view, and not the angle of incidence, the details of the scattering (including effects of surface irregularity) are common mode and drop out.  The peak-to-peak variation in the electrode surface height of roughly $\sim100$ nm (due to the roughness of the electroplated gold) must be added to the overall uncertainty in the fit.  The lateral distance calibration is completed by scattering the beam off the trap surface and measuring its position on the EMCCD camera image relative to the trap electrodes.  The ion height uncertainty is approximately 1 $\mu$m, while the lateral distance uncertainty is $\approx3\,\mu$m.

The rf trapping potential is generated by driving a critically-coupled cryogenic LC resonant circuit made from a 315 nH printed-circuit-board-based toroidal inductor similar to designs presented in Ref.~\cite{Brandl2016} shunted by a low-loss surface-mount ceramic capacitor~\cite{Todaro2020}.  The resonator has a loaded $Q$ of 169 at 4 K and provides a voltage step-up of 18.5.  We drive the resonator with {$\approx3$~mW} of rf power to provide the trapping potential.  

\begin{figure}[t]
\includegraphics[width=0.48\textwidth]{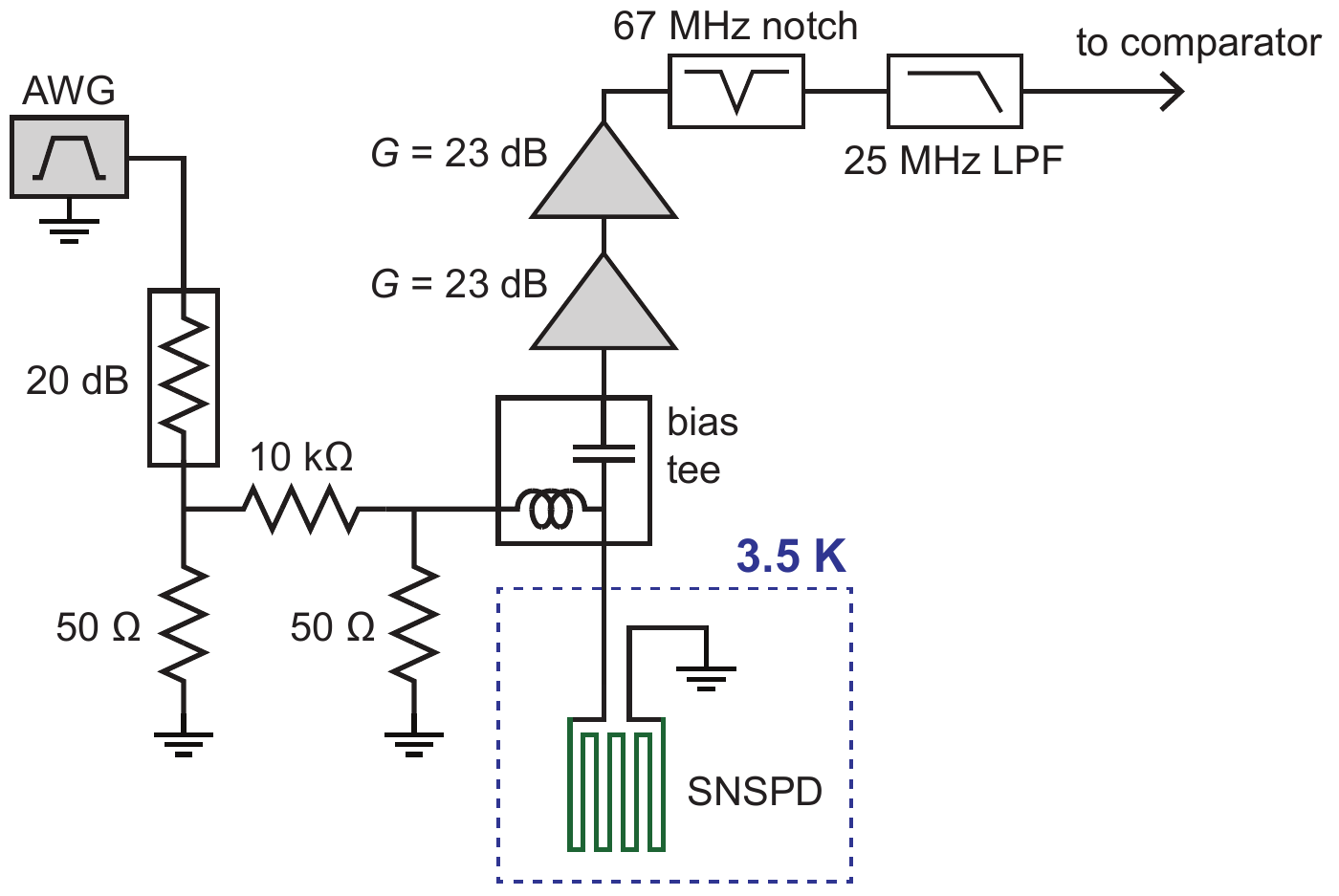}
\centering
\caption{SNSPD bias and readout electronics. See text for description.  AWG: arbitrary waveform generator; LPF: low pass filter; $G$: amplifier gain.}
\label{fig:electronics}
\end{figure} 

The SNSPD room temperature bias and readout electronics, shown in Fig. \ref{fig:electronics},  consist of an attenuated arbitrary waveform generator (AWG) for driving bias current pulses through the SNSPD and a low-noise amplifier chain for the SNSPD output signal.  These are coupled to the low-frequency and high-frequency ports, respectively, of a bias tee whose combined port is connected to the SNSPD.  The bias current pulses are ramped on and off smoothly over 11 $\mu$s to avoid ringing due to the finite bandwidth of the bias tee.  The amplifier chain provides 46 dB of gain between 1 MHz and 1 GHz.  Parasitic coupling between the trap rf electrodes and the SNSPD gives rise to a signal on the SNSPD output at the trap rf frequency whose amplitude is several times larger than the amplitude of the SNSPD pulses.  This rf pickup is removed using a 7th-order Bessel notch filter after the output amplifiers, which provides 40 dB insertion loss at the trap rf frequency, with a 3 dB bandwidth of 14 MHz.  A final dissipative Gaussian lowpass filter with $f_{\mathrm{3dB}}=25$ MHz gives an additional 18 dB insertion loss at the trap rf frequency, while also reducing the noise bandwidth.  This improves pulse discrimination, which is carried out using a high-speed Schmitt-trigger comparator.  The digital pulses from the comparator are then counted and timestamped with 1~ns resolution by the ARTIQ experimental control system~\cite{Bourdeauducq2018}.

The cryostat was operated with the ion trap at a nominal temperature as low as 3.45 K.  The SNSPD continued to operate in the presence of the trap rf with reduced $I_m$, and correspondingly reduced SDE, up to a trap temperature of 3.65 K.  The data for this paper were taken over the course of several months, during which time the base temperature of the cryostat drifted slowly upward by $\approx120$ mK, from 3.45 K to 3.57 K.  As a result, the value of $I_m$ for the data in Fig.~\ref{fig:ibias} is $\approx5.3\,\mu$A, while the value of $I_m$ (at the same rf amplitude) for the data taken to calibrate the SNSPD SDE (which were taken several months later, at a higher cryostat temperature) is $\approx4.8\,\mu$A.  

\subsection{SNSPD pulses and induced rf currents}

\begin{figure}[t]
\includegraphics[width=0.48\textwidth]{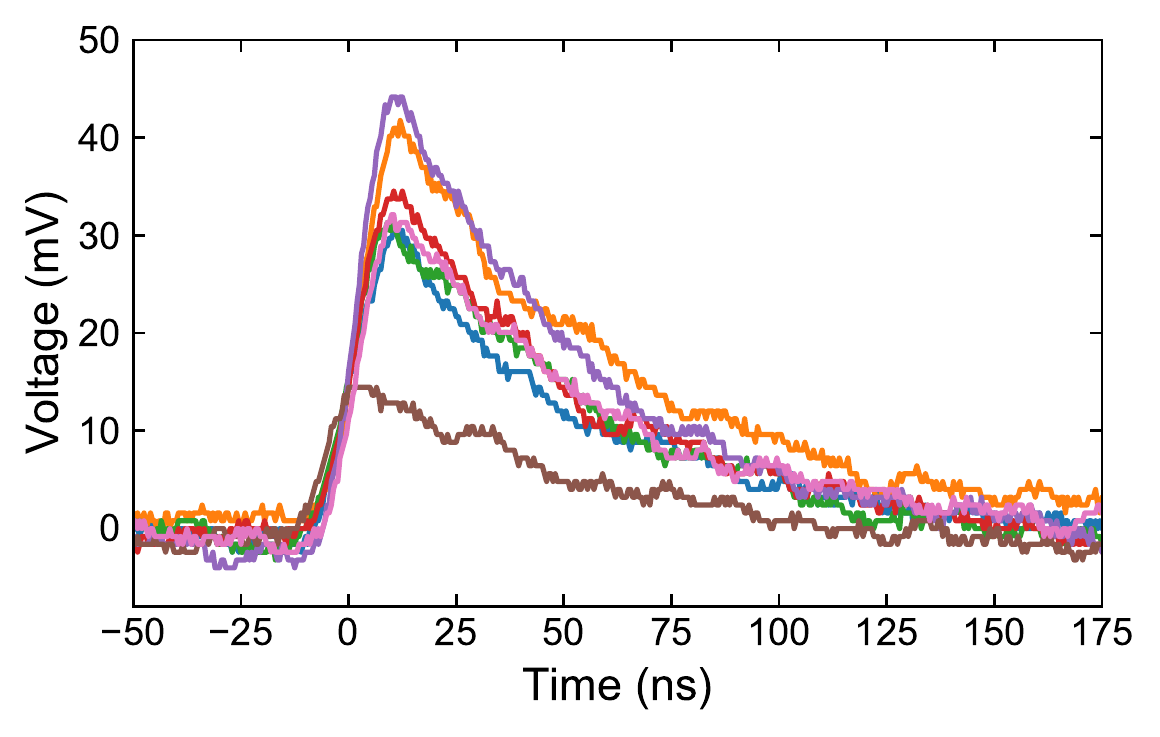}
\centering
\caption{SNSPD output pulses.  We plot seven single-shot SNSPD output pulses, measured after amplification and filtering.  The slowed $\sim$10 ns rise time (leading to increased timing jitter) and variable pulse height are evident.}
\label{fig:pulses}
\end{figure}

The height of the voltage pulse from an SNSPD is given by the product of the shunting impedance (here, the 50 $\Omega$ input impedance of the first-stage output amplifier seen in Fig.~\ref{fig:electronics}) and the instantaneous bias current, which is typically constant in most SNSPD applications.  When the trap rf is off and the filters are removed from the output chain, the output pulses from the SNSPD exhibit uniform pulse heights with $\sim$1~ns rise times.  With the trap rf on, the amplified, filtered SNSPD output pulses exhibit longer rise times of $\sim$10~ns due to filtering, and the pulse heights vary by a factor of up to 3 from pulse to pulse, as seen in Fig.~\ref{fig:pulses}.  We attribute the variation in pulse heights to induced rf currents in the SNSPD, which modulate the instantaneous bias current at the trap rf frequency~\cite{Slichter2017}.  As seen in Fig.~\ref{fig:ibias} in the main text, these currents also affect the count rate and bias parameters of the SNSPD.  

\begin{figure*}[tb]
\includegraphics[width=0.95\textwidth]{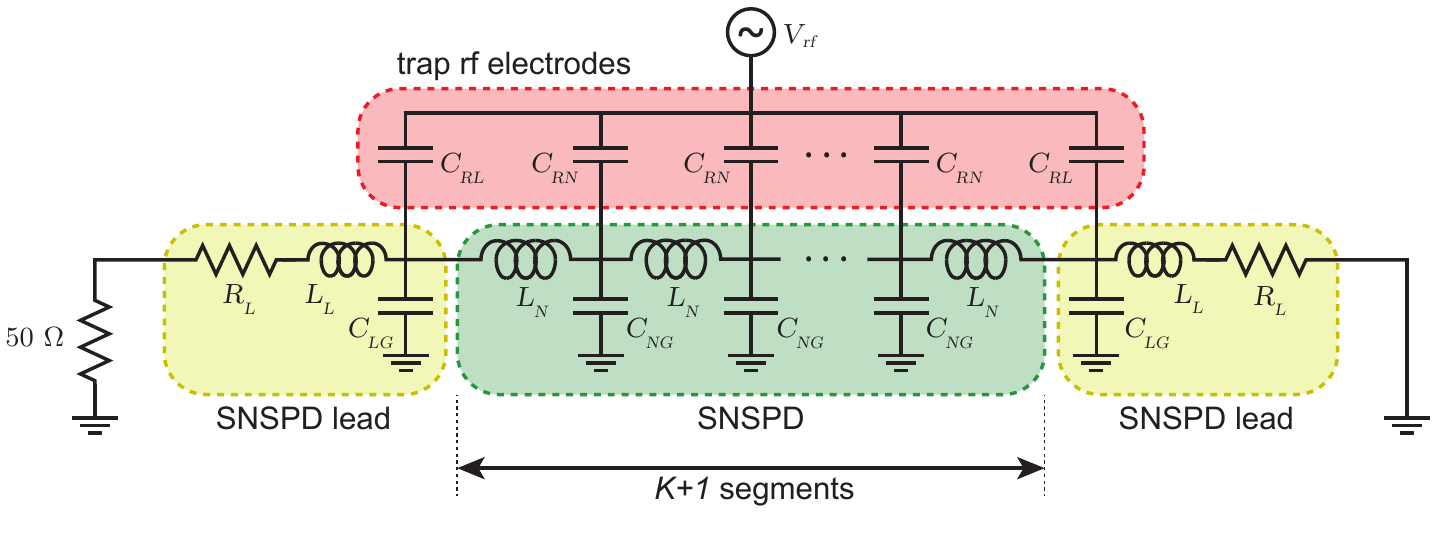}
\centering
\caption{Circuit model for induced rf currents in the SNSPD.  The SNSPD (green) is modeled as a one-dimensional LC transmission line with additional capacitive coupling to the trap rf electrodes (red).  The SNSPD leads (yellow) also have capacitive coupling to the trap rf electrodes, as well as series inductance and resistance.  One lead is grounded off-chip, while the other sees a 50 $\Omega$ impedance to ground (see Fig.~\ref{fig:electronics}).  We solve for the currents in the nanowire inductor segments to determine $I_\mathrm{rf}(k,t)$.}
\label{fig:rfmodel}
\end{figure*}

We can study these induced currents theoretically to gain insight into SNSPD performance in the presence of rf.  Figure~\ref{fig:rfmodel} shows a diagram of the circuit model used to simulate induced rf currents.  For simplicity, we model the nanowire (shown in green) as a lumped-element transmission line composed of $K+1$ segments (indexed by integers $k\in[0,K]$), with series inductance $L_N$ and capacitance to ground $C_{NG}$ per segment (the results are essentially unchanged when the value of $C_{NG}$ is increased or decreased by a factor of 10 from the calculated value).  We also consider the capacitance $C_{RN}$ per segment to the trap rf electrode, shown in red. This circuit approximation is valid because the very large kinetic inductance of the nanowire dominates any geometrical mutual inductances between neighboring nanowire segments in the meander.  We also ignore the capacitance between nanowire segments.  This capacitance modifies the effective speed of signal propagation in the nanowire, but does not play a meaningful role in the nanowire's response to the trap rf (at $\omega_\mathrm{rf}$, this capacitance just acts as a high-impedance shunt in parallel with $L_N$).  The mutual inductance between the nanowire and the trap rf electrodes is calculated to be too small to give rise to appreciable induced rf currents and is therefore not included in the model.

The induced rf current through the $k$th inductor is $I_\mathrm{rf}(k,t)$ and is in general dependent on $k$ and time $t$.  We adopt the sign convention for $I_\mathrm{rf}(k,t)$ that currents in the nanowire flowing to the right (left) in Figure~\ref{fig:rfmodel} are positive (negative).  When the trap rf is on, $I_{m}$ will be reduced by $\max_{\{k\},t}|I_\mathrm{rf}(k,t)|$; experimentally, this reduction is $\approx3.6\,\mu$A (see Fig.~\ref{fig:ibias} in the main text).  Bias currents larger than $I_{m}$ will cause the critical current to be exceeded at the location of maximum $|I_\mathrm{rf}|$ once every rf cycle, giving a dark count each time.  

The total nanowire inductance was determined from the SNSPD pulse decay time (without filters) to be 2.2 $\mu$H, and capacitances were estimated from electrostatic finite-element simulations of the trap.  The capacitance $C_{RN}$ between a segment of the nanowire and the trap rf electrode is weakly dependent on $k$ ($<15$ \% variation over all values of $k$) and is symmetric about $k=K/2$; for simplicity, we treat the $C_{RN}$ as independent of $k$.  The gold leads which contact the nanowire at its ends also have a capacitance $C_{RL}$ to the trap rf electrodes, with inductance $L_L\approx5$ nH in series with resistance  $R_L\approx5\,\Omega$ to the bias circuit (grounded on one lead, 50 $\Omega$ to ground on the other, as seen in Fig.~\ref{fig:electronics}).  The inductive impedance $Z_{L_N}=i\omega_\mathrm{rf}L_N$ is much smaller than the capacitive impedances $Z_{C_{NG}}=(i\omega_\mathrm{rf}C_{NG})^{-1}$ and $Z_{C_{RN}}=(i\omega_\mathrm{rf}C_{RN})^{-1}$ (we estimate $|Z_{C_{RN}}|\sim10^6\times |Z_{L_N}|$ and $|Z_{C_{NG}}|\sim10^4\times |Z_{L_N}|$ at $\omega_\mathrm{rf}$), and the impedance of the SNSPD leads $Z_\mathrm{lead}$ in series with the lead termination impedances is small compared to the total impedance of the nanowire $\approx (K+1)Z_{L_N}$ (again using $|Z_{L_N}|\ll |Z_{C_{RN}}|,\, |Z_{C_{NG}}|$).

The capacitive coupling $C_{RL}$ between the trap rf electrodes and the SNSPD leads induces an oscillating rf voltage at each end of the nanowire.  If the termination impedances were the same for both leads, these voltages would be the same by symmetry, and the voltage difference across the nanowire would be zero.  The asymmetric lead termination impedances make these voltages asymmetric, giving rise to a differential voltage across the nanowire.  This leads to a spatially uniform ($k$-independent) current through the nanowire, in phase with the trap rf voltage.  The magnitude of this current increases with $C_{RL}$ and with the asymmetry in termination impedances and decreases linearly with the total nanowire inductance.  

In addition, the direct capacitive coupling $C_{RN}$ between the trap rf electrodes and the nanowire induces a spatially varying ($k$-dependent) rf current in the nanowire.  The amplitude of this induced current varies linearly with $k$; the amplitudes at $k=0$ and $k=K$ have equal magnitude but opposite sign, and the amplitude at $k=K/2$ is zero (this relies on the fact that the nanowire impedance is much larger than the lead termination impedances).  The magnitude of the spatially varing induced current scales linearly with $C_{RN}$ and is independent of $L_N$ and $C_{NG}$.  The behavior can be qualitatively understood by symmetry arguments; if the lead termination impedances were equal, then by the symmetry of the circuit $I_\mathrm{rf}(k,t) = -I_\mathrm{rf}(K-k,t)$, and $I_\mathrm{rf}(K/2,t)=0$.  This spatially varying induced current is 90 degrees out of phase with the trap rf drive.  The unequal lead termination impedances mean that an additional spatially uniform ($k$-independent) current is induced by this capacitive coupling, which is in phase with the trap rf drive.  For our device parameters, the magnitude of this current is smaller than both the spatially uniform current due to the leads and the spatially varying current due to the direct nanowire coupling.  

The total induced current $I_\mathrm{rf}(k,t)$ in the nanowire is thus a sum of the two spatially uniform ($k$-independent) currents in phase with the trap rf drive, and one spatially varying ($k$-dependent) current 90 degrees out of phase with the trap rf drive.  For the parameters of our device, the spatially-varying induced current is dominant.  

The circuit model explains why attempts to increase $I_{m}$ and/or SDE by applying an rf ``cancellation'' bias current to the SNSPD, as demonstrated in Ref.~\cite{Slichter2017}, were not successful; such a cancellation tone is spatially uniform in the nanowire, and so cannot cancel the spatially-varying portion of $I_\mathrm{rf}(k,t)$, which is the dominant component of the induced rf current in this device.  For future devices, spatially-varying induced currents can be minimized by reducing $C_{RN}$ through capacitive shielding of the SNSPD.  Spatially-uniform induced currents can be minimized passively by choosing symmetric termination impedances for the SNSPD, by reducing $C_{RL}$, and by increasing the SNSPD inductance; active cancellation of such currents through an externally applied drive tone is also possible but adds experimental complexity.  Taken together, these improvements should provide substantial reductions in induced rf currents, enabling operation with higher SDE and at higher rf amplitudes, as required for heavier ion species.  Improved SNSPD materials and fabrication can provide larger ``plateau'' regions, where the SDE is relatively insensitive to bias current, further improving tolerance of induced rf currents.  

The model curve shown as a blue line in Fig.~\ref{fig:ibias} is a fit to the measured counts with rf on, assuming that the bias current is being modulated at the rf frequency $\omega_\mathrm{rf}$ with the spatial dependence described by the circuit model (specifically, Eq.~\ref{eq:pickup}).  The measured bright counts with rf off are used to estimate the instantaneous count rate for a given instantaneous bias current, which is then averaged over the rf cycle and over the full range of positions $k$ along the nanowire.  We do not account for varying photon absorption rates in the nanowire at different $k$ due to variations in the polarization, incidence angle, or local intensity of fluorescence photons at the nanowire, nor do we account for spatial variation in detection efficiency, due for example to nanowire constrictions or fabrication defects.  At very low bias currents, some output pulses may go uncounted because their maximum voltage is below the comparator threshold, so we expect the model fit to be slightly higher than the observed counts in this regime.  We use the following expression for the magnitude of the time-dependent induced rf current:

\begin{equation}
\label{eq:pickup}
I_\mathrm{rf}(k,t)=I_0\sin(\omega_\mathrm{rf}t) + I_1\left(\frac{k-K/2}{K/2}\right)\cos(\omega_\mathrm{rf}t)\,.
\end{equation}
\noindent
Here $I_0$ is the amplitude of the spatially-uniform induced current, and $I_1$ is the maximum amplitude of the spatially-varying induced current.  This expression captures the essential features of the induced rf currents as seen in the circuit model described above.  The fit yields $I_0=0.9(1)\,\mu$A and $I_1=3.5(2)\,\mu$A with $K=40$, which gives the blue curve shown in Fig.~\ref{fig:ibias}.  We constrain the fit such that the maximum $|I_\mathrm{rf}|$ (at $k=0$ and $k=K$) should be equal to the observed $I_m$ reduction of $\approx3.6\,\mu$A.

We note that the $C_{RN}$ values required to achieve a maximum $|I_\mathrm{rf}|$ of $\approx3.6\,\mu$A are about 1.5 times larger than the values determined from finite element simulations; the source of this discrepancy is not clear, but may be related to simplifying assumptions in the finite element simulation.

\subsection{Readout fidelity}

We define the bright error rate $\epsilon_{b}$ as the fraction of trials prepared in the bright state that are read out as dark, and the dark error rate $\epsilon_{d}$ as the fraction of trials prepared in the dark state that are read out as bright.  The readout fidelity is then defined as

\begin{equation}
\mathcal{F}=1-\frac{\epsilon_{b}+\epsilon_{d}}{2}\,.
\end{equation}

\noindent
For the thresholding method, $\epsilon_{b}$ is the fraction of the counts below the threshold when prepared in the bright state, and $\epsilon_{d}$ is the fraction of counts above the threshold when prepared in the dark state.  The threshold is chosen to maximize the readout fidelity as defined above.  

The adaptive Bayesian method is related to methods for improving readout fidelity by photon time-of-arrival analysis first discussed in detail in Ref.~\cite{Langer2006}.  It is similar to the Bayesian method demonstrated in Ref.~\cite{Myerson2008}, except that we have included the effects of both depumping (bright to dark transitions) and repumping (dark to bright transitions).  Unlike Ref.~\cite{Crain2019}, which uses only the arrival time of the first photon to estimate the state, we use multiple photon counts until a desired Bayesian confidence level is reached, as described below.

In this work, the Bayesian analysis is carried out entirely in post-processing on a computer for simplicity.  However, we note that it is possible to perform the necessary calculations in real time on the embedded processor of a suitable experimental control system~\cite{Bourdeauducq2018}.  This would be important for applications such as quantum error correction, where subsequent algorithmic steps are conditional on the qubit measurement outcome.  

We divide each readout trial into $N$ time bins of length $t_0$ (in our analysis, we use $t_0=1\,\mu$s) and determine the number of photon counts in each time bin, yielding a set of time bin counts $\{n_i\}$, where $i\in\{1,N\}$.  We assume that the number of counts in a given bin is Poisson distributed, with means $\gamma_bt_0$ and $\gamma_dt_0$ for the bright and dark states, respectively ($\gamma_b$ and $\gamma_d$ are the mean photon count rates for the bright and dark states, respectively).  We also assume repumping and depumping rates $\gamma_{rp}$ and $\gamma_{dp}$. The four rates $\gamma_b$, $\gamma_d$, $\gamma_{dp}$, and $\gamma_{rp}$ must be determined experimentally using independent calibration data.  We find $\gamma_b$ and $\gamma_d$ from the mean bright and dark state photon counts versus time by fitting the corresponding histogram peaks to Poissonian distributions for different readout durations.  We extract $\gamma_{dp}$ and $\gamma_{rp}$ by preparing the nominal bright and dark states and measuring the decrease in bright count rate and increase in dark count rate, respectively, of the instantaneous average count rate as a function of the time since the start of the readout.  Both sets of calibrations account for imperfect state preparation.  We measure $\gamma_b=162.50(3)\,\mathrm{ms}^{-1}$, $\gamma_d=5.095(5)\,\mathrm{ms}^{-1}$, $\gamma_{dp}=0.020(3)\,\mathrm{ms}^{-1}$, and $\gamma_{rp}=0.0120(7)\,\mathrm{ms}^{-1}$.  Variation of any of the four rates at the quoted uncertainty level has no discernible effect on the resulting Bayesian readout fidelity.  Because at least two photons must be scattered to repump the shelved state $\ket{aux}$ to the bright state $\ket{\downarrow}$, repumping involves a dwell time in one of several possible non-fluorescing hyperfine states, and thus the repumping process will not be characterized by a single time-independent rate parameter $\gamma_{rp}$~\cite{Langer2006}.  However, in the limit of only a few percent of the population in $\ket{aux}$ being repumped to $\ket{\downarrow}$, the approximation of a single rate parameter $\gamma_{rp}$ is reasonable, as confirmed by experimental data.  

We take $t_0\ll1/\gamma_{rp}, 1/\gamma_{dp}$, equivalent to the statement that depumping and repumping events are rare for any given time bin; this assumption is well supported by the measured values given above.  This allows us to make the simplifying approximation that any depumping or repumping event occurs instantaneously in between time bins, with probabilities of $\gamma_{dp}t_0$ and $\gamma_{rp}t_0$, respectively.  This approximation has negligible impact on the results of the calculation. We make a Bayesian estimate of the probability $P_i^b$ and $P_i^d=1-P_i^b$ that the ion is bright or dark, respectively, based on analysis of the data up to the end of the $i$th time bin.  The calculation of these probabilities is recursive.  We begin by assuming a uniform prior, that is, equal probability for determining that the ion is in each state:
\begin{eqnarray}
P_0^b &=& 0.5\, , \nonumber\\
P_0^d &=& 0.5 \,.
\end{eqnarray}
At the end of the $i$th time bin, the (non-normalized) posterior probabilities $p_i^b$ and $p_i^d$ are given by
\begin{eqnarray}
p_i^b &=& \left[(1-\gamma_{dp}t_0)P_{i-1}^b + \gamma_{rp}t_0 P_{i-1}^d \right] f(n_i,\gamma_b)\, , \nonumber \\
p_i^d &=& \left[(1-\gamma_{rp}t_0)P_{i-1}^d + \gamma_{dp}t_0 P_{i-1}^b \right] f(n_i,\gamma_d) \, , 
\end{eqnarray}

\noindent
where $f(n,\gamma) = \frac{1}{n!}(\gamma t_0)^n e^{-\gamma t_0}$ is the Poisson distribution with mean $\gamma t_0$.  The term in brackets in the first (second) expression is the prior, the probability that the ion was bright (dark) at the start of $i$th time bin, including the effects of depumping and repumping (taken to occur instantaneously between the time bins $i-1$ and $i$ as described above).  This is multiplied by the likelihood of observing $n_i$ counts for such a bright (dark) ion during the $i$th time bin.  The resulting normalized posterior probabilities that the ion is bright or dark after the $i$th bin are calculated as
\begin{eqnarray}
P_i^b &=& \frac{p_i^b}{p_i^b + p_i^d}\, , \nonumber \\
P_i^d &=& \frac{p_i^d}{p_i^b + p_i^d}\, .
\end{eqnarray}
The Bayesian readout method is made ``adaptive'' by monitoring the probabilities $P_i^b$ and $P_i^d$, stopping the readout and declaring the state to have been determined when one of them reaches a certain level.  The different data points for the Bayesian readout fidelity in Fig.~\ref{fig:hist}(b) correspond to different values of this state determination confidence level, logarithmically spaced between 0.9 and 0.9999.  Note that this confidence level for state determination is distinct from the readout fidelity.  Because the value of $i$ at which the desired confidence level is reached will depend on the measurement record $\{n_i\}$ for each trial, the readout duration will vary from shot to shot, and thus the horizontal axis in Fig.~\ref{fig:hist}(b) shows the average readout duration over $10^5$ trials each of preparing the bright and dark states.  The asymmetry between $\gamma_b$ and $\gamma_d$ means that a bright ion will reach the state determination threshold faster on average than a dark ion.  For the highest readout fidelity of 0.9991(1), the state-averaged mean readout duration was 46 $\mu$s; for the bright state the mean readout duration was 25 $\mu$s with $\epsilon_b=6.1\times10^{-4}$, while for the dark state it was 67 $\mu$s with $\epsilon_d=11.9\times10^{-4}$.  This asymmetry in readout duration and achievable readout fidelity can be harnessed to improve readout performance for quantum algorithms whose most likely output is known (such as syndrome measurements in quantum error correcting codes) by adding qubit control pulses to map the most likely output state to the bright state of the ion.

The achievable readout fidelity depends on the atomic level structure of the ion species; ions with metastable $D$ or $F$ states that can be used for shelving, such as Ca$^+$, Sr$^+$, Ba$^+$, or Yb$^+$, can potentially achieve higher readout fidelities than ions that lack these states, such as $^9\mathrm{Be}^+$.  For example, assuming the same overall photon detection efficiency reported here, the mean readout error in $^{40}\mathrm{Ca}^+$ could be as low as $\approx2\times10^{-5}$, limited by the lifetime of the $D_{5/2}$ state used for shelving~\cite{Myerson2008}.

\subsection{Photon arrival time correlations}

\begin{figure}[tb]
\includegraphics[width=0.48\textwidth]{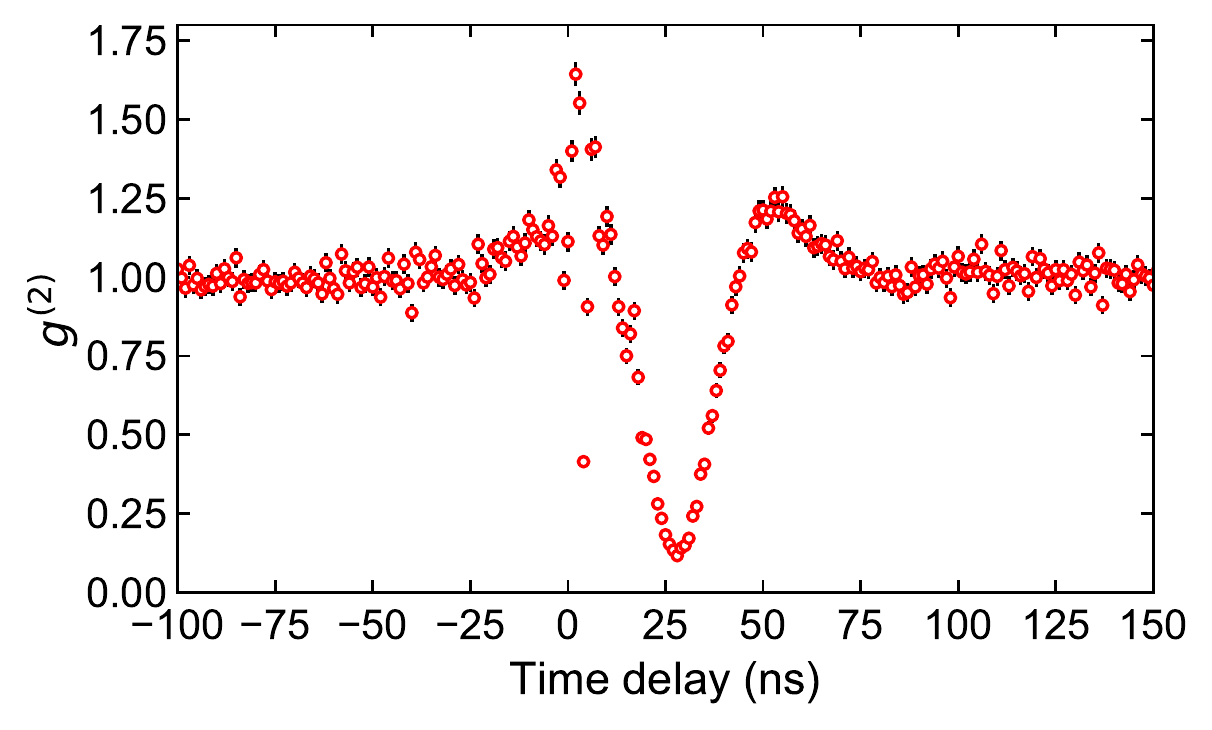}
\centering
\caption{SNSPD and PMT photon correlations.  We plot the $g^{(2)}$ correlation function between photon arrival times measured by the SNSPD and the PMT when both are used to count fluorescence photons during the same readout period.  No background correction is performed.  The minimum value of $g^{(2)}$ is shifted to 28 ns delay, rather than the expected 0 ns, due to the larger signal propagation delays in the SNSPD amplification/comparator chain relative to the PMT amplification/comparator chain.  Data close to 0 ns delay are corrupted due to electrical crosstalk in the time-tagging electronics.  Black error bars (smaller than the symbols for many points) represent 68 \% confidence intervals.}
\label{fig:g2}
\end{figure}

We can count fluorescence photons with both the SNSPD and the PMT simultaneously during the same detection period.  By timestamping the photons with 1 ns resolution, we are able to compare photon arrival times at the two detectors and study the statistics of the counted photons.  Since the photons come from a single atom, we anticipate that they should be anti-bunched.  We plot the $g^{(2)}$ correlation function of the photon arrival times for the SNSPD and PMT in Fig.~\ref{fig:g2}.  The corruption of the data near zero time delay is an experimental artifact due to electrical crosstalk in our time-tagging electronics.  The minimum in $g^{(2)}$ correlation at 28 ns delay is due to the anti-bunching of ion fluorescence photons.  This minimum is shifted to 28 ns delay from the nominal value of 0 ns due to larger signal propagation delays in the SNSPD amplification/comparator chain (primarily due to the filters and additional cable length) relative to the PMT amplification/comparator chain.  The plot shows the raw data, where no detector background count subtraction has been performed.  The minimum value of $g^{(2)}$ is limited by the PMT and SNSPD count rates due to stray laser light, which are considerably higher than the intrinsic dark count rates of either detector.

\subsection{Ion fluorescence emission and detector SDE}

The SDE of the SNSPD is given by the expression $\mathrm{SDE}=\mathrm{AP}\times\mathrm{IDE}$, where AP is the absorption probability and IDE is the intrinsic detection efficiency. The absorption probability is the probability that a photon incident on the SNSPD is absorbed in the nanowire, and not reflected or absorbed in another location such as the Si substrate. The intrisic detection efficiency is the probability that an absorbed photon is converted to an electrical output pulse~\cite{Engel2015}.  The IDE depends on the bias current, among other factors, while the AP is independent of the bias current.  The presence of a plateau in the SDE versus bias current (as seen in the data with rf off in Fig.~\ref{fig:ibias} in the main text) is generally accepted as evidence for saturation of the IDE close to unity~\cite{Baek2011}, although definitive experiments to measure the IDE in these situations have not been carried out to date.  SNSPDs made from amorphous MoSi (such as in this work) typically show large plateaus, especially at shorter wavelengths~\cite{Verma2015, Wollman2017, Caloz2017}.  Since the signal-to-noise degradation of an SNSPD output pulse due to the amplifiers and filters is negligible (assuming sufficiently large bias current), the SDE should be equal to the AP when the bias current is in the plateau region.  

\begin{figure}[t]
\includegraphics[width=0.48\textwidth]{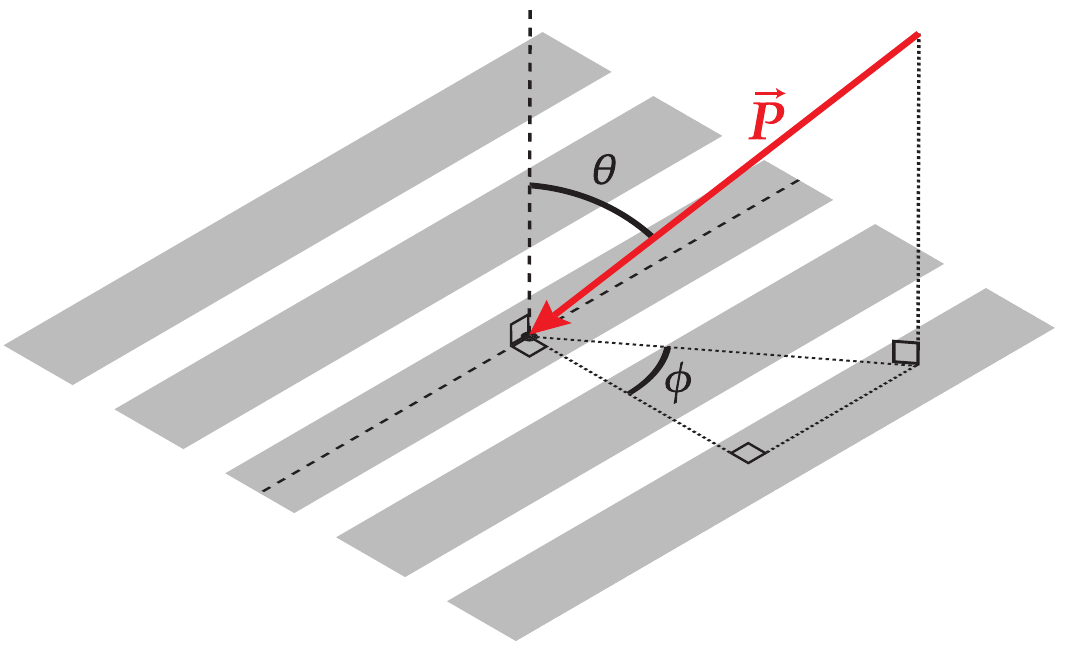}
\centering
\caption{Geometry definition for AP calculation.  The Poynting vector $\vec{P}$ (red) of incoming radiation is related to the geometry of the nanowires (grey) with the polar angle $\theta$ and azimuthal angle $\phi$ as shown.  The nanowires in our trap are oriented parallel to the trap axis.  }
\label{fig:poynting}
\end{figure}

\begin{figure*}[tb]
\includegraphics[width=0.85\textwidth]{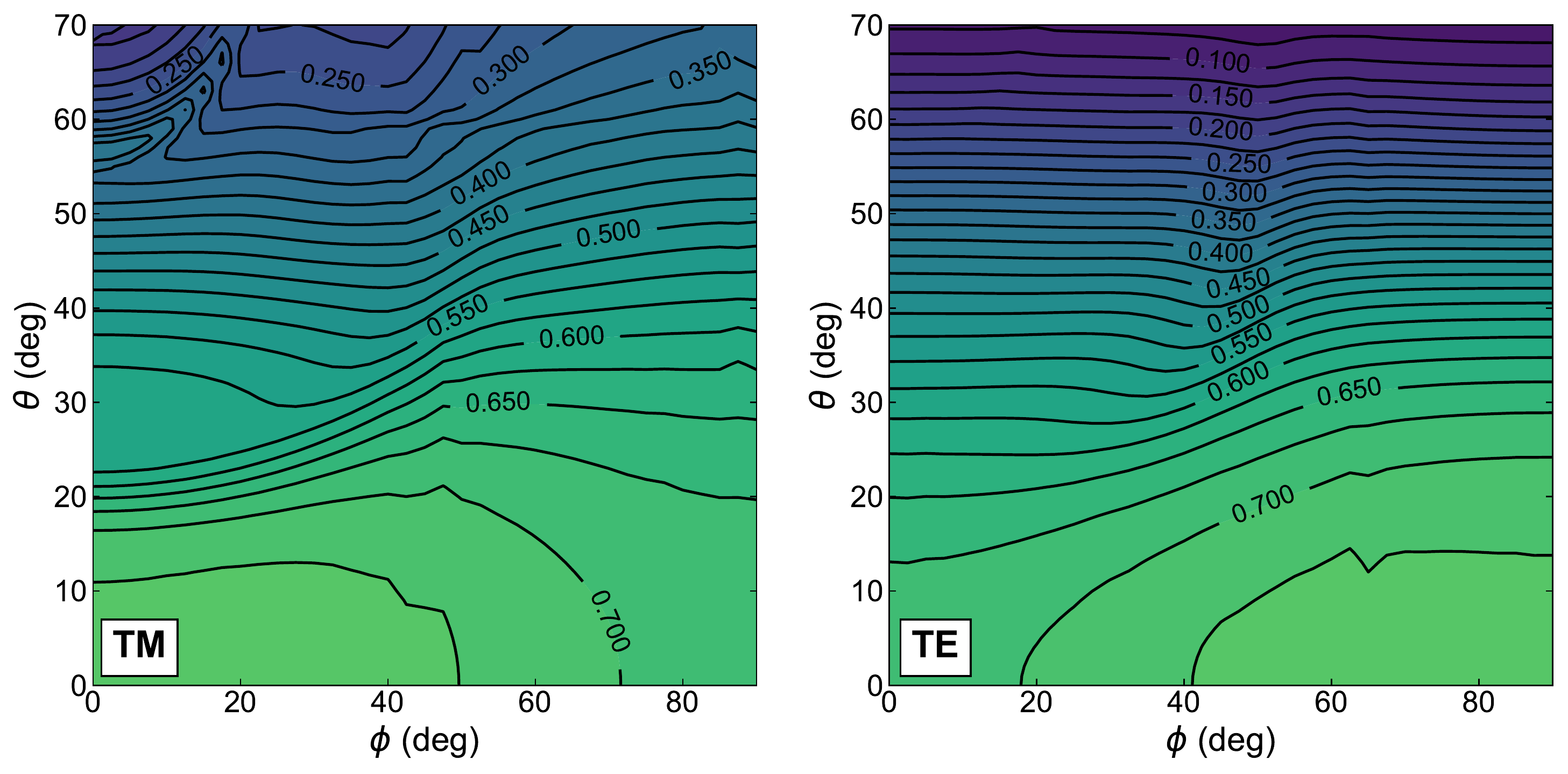}
\centering
\caption{Calculated SNSPD photon absorption probability.  Contour plot of angle-dependent photon absorption probability (AP) from finite element analysis calculations for TM (left) and TE (right) polarizations.  Contours are spaced at intervals of 0.025.}
\label{fig:sde_thy}
\end{figure*}

 We calculate the AP, including its dependence on photon incidence angle and polarization, using finite element analysis in COMSOL~\cite{disclaimer}.  Plane wave radiation incident on the SNSPD is simulated for a grid of $(\theta,\phi)$ values, where $\theta$ is the polar angle and $\phi$ is the azimuthal angle of the Poynting vector of incident radiation as defined in Fig.~\ref{fig:poynting}.  For efficient computation, we simulate an infinite periodic array of nanowires by tiling a ``unit cell'' geometry in the two in-plane dimensions with periodic (Floquet) boundary conditions.  We perform simulations for both TE and TM polarizations, where the electric field or the magnetic field, respectively, of the incoming radiation lies in the plane of the nanowires.  The calculated angle-dependent AP for both TE and TM polarizations is shown in Fig.~\ref{fig:sde_thy}. Arbitrary incident polarizations can be represented as linear combinations of TE and TM polarizations with complex field amplitudes.  The contribution of the TE and TM components to the average AP are weighted by their intensities (modulus squared of the complex electric field amplitude).   For normally incident photons, the AP averaged over the azimuthal angle $\phi$ is 72~\%, in agreement with RCWA simulations.  As stated above, for bias currents on the SDE plateau, close to the nanowire switching current, we take $\mathrm{IDE}\approx1$, meaning the SDE is approximately equal to the AP.  

Because the fluorescence photons from the ion have $\sigma^{-}$ polarization relative to $\vec{B}_0$, the photon flux from the ion in the far field is not isotropic.  Instead, the intensity varies with the angle $\theta_q$ between the fluorescence Poynting vector and the quantization axis as $1+\cos^2\theta_q$~\cite{Jackson1999,Budker2008}.  This must be accounted for in calculations of the fraction of fluorescence photons incident on the SNSPD from the ion. 

We calculate the theory curves in Fig.~\ref{fig:crosstalk} by assuming that the SNSPD count rate $\kappa$ when the atomic transition is driven with a saturation parameter $s\gg1$ is given by the integral

\begin{equation}
\label{eq:kappa}
\kappa = \frac{\Gamma}{2}\int_{\Omega_{det}}d\Omega\,\frac{3}{16\pi}\big[1+\cos^2\theta_q\big]\times\mathrm{IDE}\times\mathrm{AP}\, \, ,
\end{equation}

\noindent
 where the integral is computed over the solid angle $\Omega_{det}$ subtended by the SNSPD as viewed from the ion position.  Here $\Gamma$ is the atomic transition linewidth.  The two theoretical curves in Fig.~\ref{fig:crosstalk} in the main text assume either an angle-independent AP (red curve) or an angle-dependent AP as described above (green curve).  Because of induced rf currents and the choice of $I_b$ below the plateau value, the IDE will be less than one, although it will be independent of ion position.  We calibrate this IDE by adjusting it so the theoretical value of $\kappa$ matches the experimental data at the left-most experimental data point.  This provides an overall scaling for the IDE and thus the SDE.

 We perform the integration in Eq.~\ref{eq:kappa} numerically by treating the SNSPD as being composed of a grid of $1\,\mu\mathrm{m}\,\times\,1\,\mu$m squares and summing the value of the integrand over these squares, accounting for the solid angle $d\Omega$ subtended by each square.  At each ion position, we use the experimentally measured ion height and lateral distance from the SNSPD to determine, for each square in the grid, the Poynting vector (and thus $\theta$ and $\phi$ for the incident photons at the detector), $\theta_q$, the polarization of the fluorescence, and the solid angle subtended by the square.  Because the detector is recessed below the top surface of the trap electrodes, it will be partially or fully obscured from the ion when the ion moves far enough away along the trap axis; however, for the range of ion positions considered in Fig.~\ref{fig:crosstalk}, the ion remains close enough to the detector that no such obscuration occurs.  
 
 The photon collection fraction of 2.0(1)~\% quoted in the main text is calculated from Eq.~\ref{eq:kappa} by setting $\mathrm{IDE}=\mathrm{AP}=1$ and dividing the resulting expression by $\Gamma/2$ to normalize for the rate at which the ion spontaneously emits fluorescence photons.  The uncertainty in this photon collection fraction is primarily due to uncertainty in the ion height.  
 
We can extrapolate from the experimentally measured SDE of 48(2)~\% to determine what the maximum SDE of the SNSPD would be in the absence of rf.  This is equivalent to determining the ratio of the IDE between rf off and rf on, since the AP is independent of the bias current.  This IDE ratio can be estimated from the data presented in Fig.~\ref{fig:ibias}, by dividing the dark-count-corrected bright count rate with rf off at the highest value of $I_b$ by the dark-count-corrected bright count rate with rf on and $I_b$ set to 0.8~$\mu$A below $I_m$.  Since the AP is constant, this is also the ratio of the SDE between rf off and rf on.  We multiply it by the SDE with rf on to yield the estimated maximum SDE with rf off of 65(5)~\%.  An uncertainty of $\pm0.2\,\mu$A in the exact value of $I_m$ (for both the data in Fig.~\ref{fig:ibias} and the saturation-based SDE calibration data) is responsible for the increased fractional uncertainty on the estimated maximum SDE without rf.  Based on finite-element simulations, the oxidation of the top 4 nm of the MoSi SNSPD material would change the average normal-incidence AP from 72 \% to 65 \%.

\end{document}